\documentclass[conference]{IEEEtran}
\ifCLASSINFOpdf
\else
\fi
\usepackage{amsmath}
\usepackage{multirow}
\usepackage{subfigure}
\usepackage{caption}
\usepackage{graphicx}
\usepackage{epsfig}
\usepackage{epstopdf}
\usepackage{cite}
\usepackage{url}
\usepackage{booktabs}
\usepackage{algorithm}
\usepackage{algpseudocode}

\makeindex
\begin{document}

\title{Real-Time Human-Robot Interaction for a Service Robot Based on 3D Human Activity Recognition and Human-mimicking Decision Mechanism}
\author{\IEEEauthorblockN{Kang Li, Jinting Wu, Xiaoguang Zhao and Min Tan}
\IEEEauthorblockA{The State Key Laboratory of Management and Control for Complex Systems,\\ Institute of Automation, Chinese Academy of Sciences,\\ University of Chinese Academy of Sciences, Beijing, China\\
Email: {likang2014@ia.ac.cn, wujinting2016@ia.ac.cn, xiaoguang.zhao@ia.ac.cn, min.tan@ia.ac.cn}}}
\maketitle
\maketitle

\begin{abstract}
  In this paper, we present a real-time Human-Robot Interaction (HRI) system for a service robot based on 3D human activity recognition and human-mimicking decision mechanism. A three-layer Long-Short-Term Memory (LSTM) network is trained to recognize human activity. The input of this network is the 3D skeleton joints information captured by Kinect V1 and the output of this network is the class label of human activity. Moreover, the human-mimicking decision mechanism is also fused into the online test process of human activity recognition, which allows the robot to instinctively decide whether to interrupt the current task according to task priority. The robot can make appropriate responses based on aforementioned analytical results. The system framework is realized on the Robot Operating System (ROS). The real-life activity interaction between our service robot and the user was conducted to demonstrate the effectiveness of developed HRI system.
\end{abstract}

\IEEEpeerreviewmaketitle

\section{Introduction}

  As robots move into our lives, the importance of enabling users to interact with them in a natural way increases \cite{Sheridan2016Human}.
  Non-verbal communication has become the most natural manner between robots and users. Therefore, the robot with the capability of understanding human body language is a trend of development in HRI system.

  In the past few years, gesture or activity recognition on RGB video processing has been widely studied and applied in various fields \cite{Waldherr2000A,Sun2009Hierarchical,Sigalas2010Gesture,Le2011Learning,Wang2013Dense}. They attempted to learn discriminative features to enhance identification performance of human activities. However, hand-crafted features only have limited discriminative power, another drawback is that the 2D image only involves limited body movement information. With the rise of deep learning techniques \cite{Lecun2015Deep}, high-level semantic features of human activities can be extracted to improve the accuracy of human activity recognition. Especially the emergence of Recurrent Neural Networks (RNN) makes sequence learning easier \cite{Du2015Hierarchical,Zhu2016Co,Song2016An,Zhang2017On}. On the other hand, 3D human activity information was also gradually applied to human activity recognition. Several researchers relied on stereo camera to obtain 3D position of human's head, hands or any other parts of body \cite{stiefelhagen2004natural,Kai2004Real,Yang2007Gesture}, Others encode the motion characteristics of an action by using depth maps from RGB-D camera \cite{Chen2016Real,Chen2015Action,Chen20163D}. Using more rich 3D human activity information makes human activity recognition much easier and more accurate.

  Unfortunately, the motivation of aforementioned approaches is that obtaining more accurate recognition results on a benchmark does not pay sufficient attention to the cost of time. And now there is no any one method can achieve 100$\%$ recognition accuracy, if the robot identification failed, corresponding wrong interactive task will go on until finished, which is time-consuming and power-hungry. Besides, if the robot is performing a task, a higher priority task is coming, the existing HRI system based on human's activity recognition will have to wait for the end of the current task, which looks very silly.

  This work demonstrates progress towards a HRI system for a service robot capable of understanding common multi-modal human activities in real time and determining whether to interrupt current interactive task cleverly. The 3D data of the skeleton joints consist of rich body movement information, which is collected from users via Microsoft Kinect and is used to train a three-layer Long-Short-Term Memory (LSTM) \cite{Lecun2015Deep} network for user's activity recognition. Our previous work \cite{Li2017Sequential} elaborated the model and learning method. In this work, we described a
  interaction switch based on filter model of attention theory, simple and efficient gesture tells the service robot to start interaction. Combined with the interaction switch, human-mimicking decision mechanism of interrupting interactive tasks was also designed. Interruption can be triggered by the user when the interaction switch turns on, human-mimicking decision -- "priority should be given to important urgent matters" will be executed. The overall HRI system was implemented into the Robot Operating System (ROS) and evaluated on a service robot. Results from human-robot interaction experiments showed that aforementioned well-designed rules enabled the robot to become more smart.

  The remainder of the paper is organized as follows. Section II describes the robot platform for HRI. Section III contains an overview of the human-robot interaction system framework and elaborates the details of method. Section IV provides the experimental results and analysis. Finally, Section V concludes this paper.

\section{Robot Platform and Configuration}

  \begin{figure}[!t]
	  \centering
	  \includegraphics[width=0.46\textwidth]{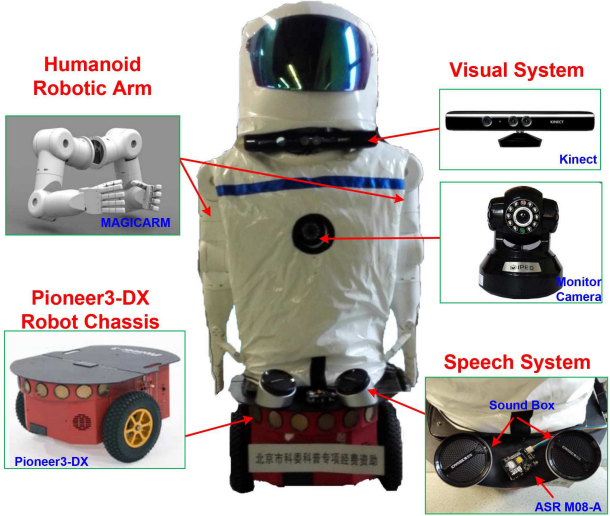}
	  \caption{Structure of service robot.}
	  \label{fig:Robot}
  \end{figure}

  Our service robot adopts a distinctive modular concept to simplify mechanical design and installation. As shown in Fig.~\ref{fig:Robot}, the service robot consists of four principal components according to function: speech system, visual system, humanoid robotic arm, and Pioneer3-DX robot chassis.
	
  1) The speech system includes an ASR M08-A module and two mini sound, which can identify speech command and broadcast the response, respectively.
	
  2) The visual system includes a Microsoft Kinect and a remote monitor camera, which is used to observe and perceive the surrounding environment. Microsoft Kinect utilizes light coding to measure depth information. The measuring range can be up to approximately 3.5~m and the depth resolution can be up to 2~mm at a distance of 2~m. The field of view (FOV) is the vertebral rectangle, the horizontal view closes to 58$^\circ$, and the resolution of the depth image is 640$\times$480 pixels. The remote monitor camera is a common webcam and mainly responsible for monitoring.

  3) In order to design friendly responses of the service robot when the user greets the robot, two humanoid robotic arms are installed on the service robot platform. The arms and hands have twelve and ten degrees of freedom, respectively. Generally, the two arms are 1.26~m long and weigh 15~kg. Each joint is a digital steering engine driven by pulse width modulation (PWM).
	
  4) The Pioneer3-DX robot chassis is the main motion component, which has one follower wheel to maintain the balance of the robot and two driven wheels to control moving. In addition, the chassis is equipped with 16 sonar detectors to perceive obstacle of environment.

  In this study, we only use part of the hardware comprising Microsoft Kinect, the robotic arms, and the Pioneer3-DX chassis for HRI.

\section{Human-Robot Interaction System}

\subsection{Overview of the framework of human-robot interaction system}

\begin{figure*}[!t]
 \centering
 \includegraphics[width=0.85\textwidth]{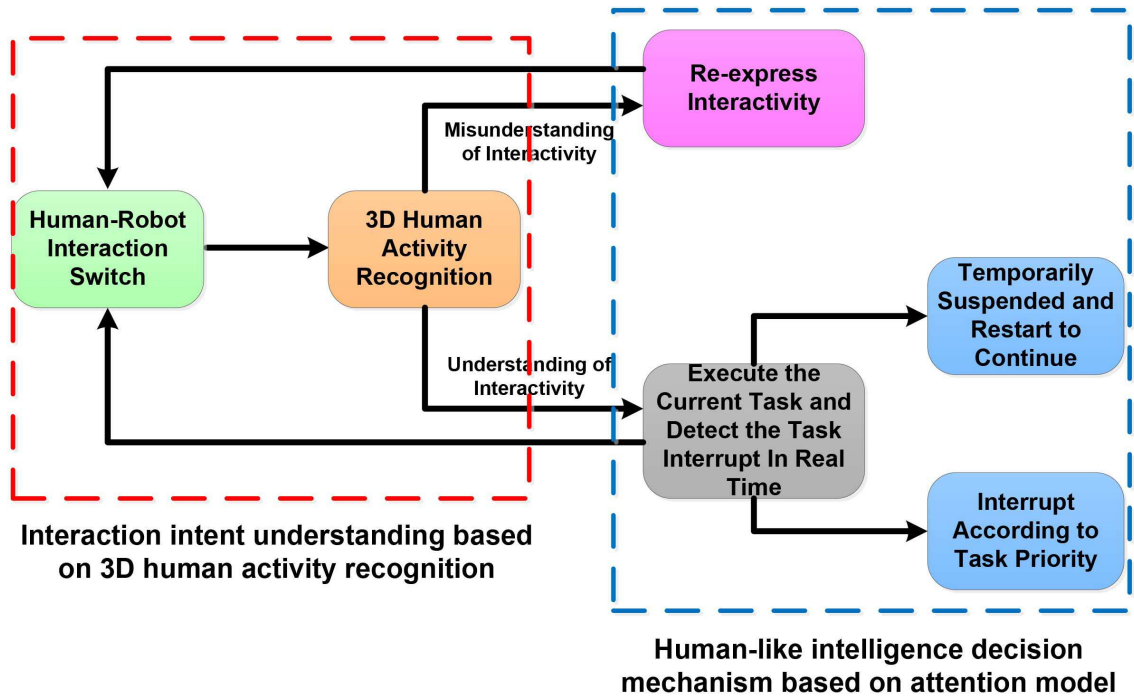}
 \caption{The framework of human-robot interaction system.}
 \label{fig:system}
\end{figure*}

Fig.~\ref{fig:system} shows the framework of human-robot interaction system. The red dashed box for understanding the interactive intent of the service robot mainly includes two modules of human-robot interaction switch and 3D human activity recognition. Human-robot interaction switch uses the RGB-D camera to obtain the human skeleton point data stream, and opens the human-robot interaction according to well-designed natural gesture. Human activity recognition module trains a three-layer LSTM recurrent neural network and uses it to infer the meaning of human activity online. Blue dashed box for the task interruption mechanism mainly includes two strategies. One is the handling method when the interactive intent is misunderstood, the other is the coping approach of sudden task disruption during the execution of current correct task. In our method, when the intent of interactivity is misunderstood, combining with the designed human-robot interaction switch and 3D human activity recognition method, it is able to interrupt the wrong task execution and re-order the robot. When the intent of interactivity is correctly understood, the tasks will be executed and the disruptions will be detected in real time. Once the interruption occurs, robot will respond interruption according to priority of the task. Such human-mimicking decision mechanism is fused into HRI system, which realizes task suspension and switching, and enhances the intelligence and flexibility of interaction.

\subsection{Human-robot interaction switch based on filter model of attention theory}

\subsubsection{The Filter Model of Cognitive Psychology Attention Theory}

In cognitive psychology, the kernel of attention is the choice analysis of information. D.E.Broadbent believed that the information from the outside world is large, while the human nervous system has a limited central processing power. In order to avoid overloading the system, some sort of filter is required to adjust it and select the lesser and more critical information to move into the advanced analysis phase. Such information will be further processed and identified and stored, while other information will not be accepted. This is the theoretical filter model \cite{Broadbent1958Perception}.

In the process of human-to-human interaction, we unconsciously notice each other's actions and filter out many less meaningful messages. In the process of human-robot interaction, in addition to the robot's ability to understand human's interactivity naturally and efficiently, it is also necessary for the robot to pay attention to human actions in real time during the mission, filter out useful information, such as people want to interrupt the current task of the intention. In our HRI system, the user can lift the left hand to attract the robot's attention. This simple and reliable skeleton-based interaction switch can make the robot filter out most of the worthless information, especially in multi-player scenarios.

\subsubsection{The Implementation of Human-Robot Interaction Switch}

\begin{figure}[!t]
 \centering
 \includegraphics[width=0.45\textwidth]{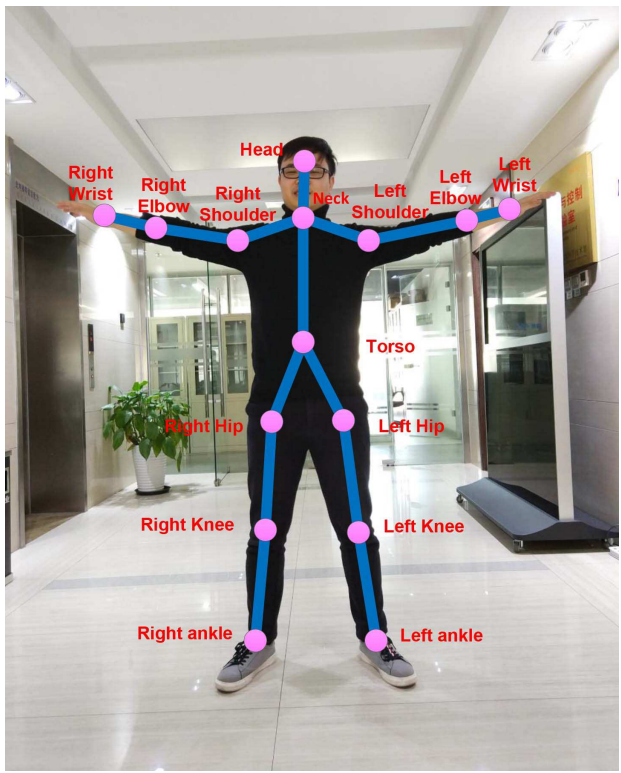}
 \caption{Human skeleton point distribution.}
 \label{fig:skeleton}
\end{figure}

\begin{figure}[!t]
 \centering
 \includegraphics[width=0.46\textwidth]{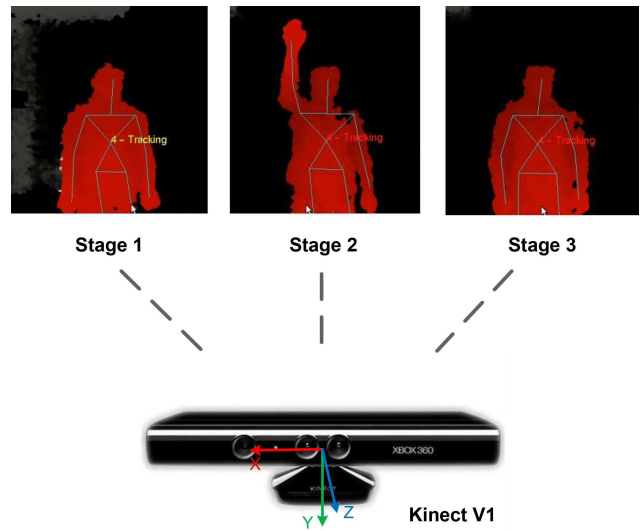}
 \caption{The implementation process of human-robot interaction switch.}
 \label{fig:stage}
\end{figure}

Human skeletal point information refers to the spatial position coordinates of the vital joint point of the human body relative to the camera coordinate system collected by the RGB-D camera. As shown in Fig.~\ref{fig:skeleton}, the human skeleton points collected by Microsoft Kinect mainly include the head, neck, torso, left wrist, left elbow, left shoulder, left hip, left knee, left ankle, right wrist, right elbow, right knee and right ankle total of 15 sets of data, and each set of data includes x, y, z coordinates of the three directions in the camera coordinate system. Therefore, the entire human skeleton point data can be expressed as a set of 45-dimensional vectors:
\begin{equation}
 \label{eq1}
 {F_n} = [{x_1},{y_1},{z_1},{x_2},{y_2},{z_2},...,{x_{15}},{y_{15}},{z_{15}}]
\end{equation}
where ${F_n}$ represents the human skeleton point data in the \emph{n}-th frame.

In the theory of attention in human's cognitive psychology, attention refers to the direction and concentration of a certain object by a mental activity. Following the characteristics of the notable object, we use the left hand gesture to turn on the interaction, which can stably and reliably attract the service robot's attention.

\begin{figure*}[htbp]
 \centering
 \includegraphics[width=0.9\textwidth]{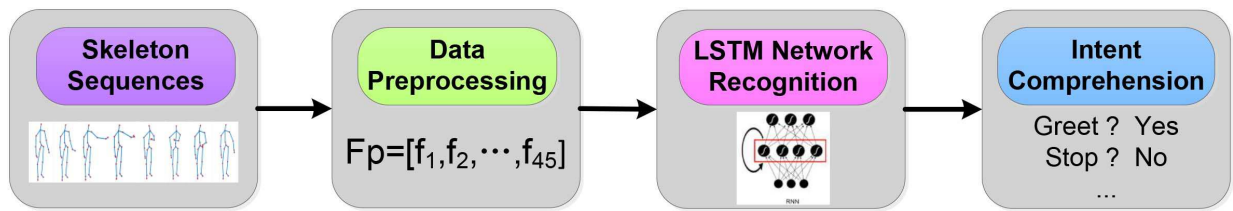}
 \caption{The basic flow of interactive intent inferring based on human activity recognition.}
 \label{fig:intent}
\end{figure*}

As shown in Fig.~\ref{fig:stage}, the designed human-robot interaction switch is divided into three stages. In the first stage, the person stood normally and both arms fell naturally. At this point, both coordinate values of the left wrist and the right wrist of the human in the y direction are greater than the coordinate value of the human torso in the y direction. Here, the flag is set to false. In the second stage, the person raised his left arm. The coordinate value of the right wrist in the y direction is still greater than the coordinate value of the human torso in the y direction, but the coordinate value of the left wrist in the y direction is smaller than the coordinate value of the human torso in the y direction. Thus, the flag is set to true. In the third stage, the person put down the left arm. According to the coordinate value of the left wrist in the y direction is greater than the coordinate value of the human torso in the y direction and the flag is true, the robot would begin to interact with the person. The person can perform a body activity in 3.5~s, the data during this time would be recorded and fed into human activity recognition module.

\subsection{Interactive intent inferring based on human activity recognition}

Fig.~\ref{fig:intent} illustrates the basic flow of interactive intent inferring based on human activity recognition. First, turning on the human-robot interaction switch to collect the skeleton data of user's activity. Then, the skeleton data will be pre-processed by affine transformation to reduce the sensitivity from differences of user's location. Finally, we adopt LSTM networks to model human activity and determine user's interactive intent. The more details of three-layers LSTM model can be found in our previous work \cite{Li2017Sequential}.

\subsection{Motion control of robotic arm and chassis}

After acquisition of user's interactive intent, the service robot responds correspondingly. Table.~\ref{tab:Interction} shows eight types of interaction between the user and the robot. Here are two motion responses we designed for the robot: the motion of the robotic arm (Interaction 1, 3, 4 and 8) and the movement of the chassis (Interaction 2, 5, 6 and 7). For robotic arm motion control, the kinematics model is established using the Denavit-Hartenberg method \cite{Hartenburg1965Kinematic}. Concrete parameters of the model can be found in our previous work \cite{Li2017Body}. For robotic chassis motion control, in order to stop the robot movement, the speed of chassis is set as 0. Moving forward and backward are both assigned fixed speed and mileage parameters, circling is corresponding to spinning. These responses will be executed when the robot understands the corresponding human activities.

\begin{table}[!ht]
	\centering
	\caption{eight common types of interaction between user and robot}\label{tab:Interction}
	\begin{tabular}{ll}
		\toprule
		Human Activities (Interactive Intents)  & Robot Responses \\
		\midrule
		1.  waving right hand (greet)           &      waving$\_$right$\_$hand \\
		2.  stretching right hand (stop)          &      stoping \\
		3.  saluting (salute)               &      saluting  \\
		4.  lifting right arm (lift right arm)  &      lifting$\_$right$\_$arm  \\
		5.  waving forwards (go back)       &       moving$\_$backwards \\
		6.  waving backwards (go ahead)                    &      moving$\_$forwards \\
		7.  drawing circle (perceive environment)  &   circling \\
		8.  waving arms around (march on the spot)      &   waving$\_$arms$\_$around \\
		\bottomrule
	\end{tabular}
\end{table}

\subsection{Human-mimicking decision mechanism}

In our HRI system, two cases of interactive task interruption are considered. One is to misunderstand the user's intent, the other is to understand the user's intent but need to interrupt current task. For the first case, the robot will carry out wrong task based on wrong interactive intent inferring. In the past HRI system, the user has to wait for the end of wrong task and restart the correct task. In our method, when the user observes that the robot misunderstand the user's intent, the current task can be suspended by the human-robot interaction switch. After the user lifts his left hand and attracts the robot's attention, the robot will stop the current task. When the user drops his left hand, the RGB-D camera will re-collect the user's body activity and re-infer the user's intent. This minor trick is consistent with human's smart decision mechanism. If others misunderstand the person's intent, the person will choose to re-express. For the second case, when the user's intent is correctly inferred, the robot will execute corresponding task. If you suddenly want to suspend the current task and re-start a higher priority task at this time, traditional interaction manner can not work. In our method, After the user lifts his left hand and draws attention of the robot, the robot will suspend the current task and record the breakpoint of the task. When the user put down his left hand, he can stop the robot using the stop command and handle his own thing, or uses the other command to control the robot. If stoping, the robot will forget the breakpoint and reset. If starting new interaction and the new command has higher priority, the new task will be opened and breakpoint will remember newest value. If starting new interaction and the new command has lower priority, the current task will be not interrupted and continued to execute. The overall algorithm has been described in Algorithm.~\ref{code:Algorithm}.

\begin{algorithm}[h]
  \caption{The procedure of human-robot interaction} \label{code:Algorithm}
  \begin{algorithmic}[1]
    \State \emph{loop}:
    \State opening human-robot interaction switch;
    \State capturing skeleton data and feeding data into LSTM network to infer user's interactive intent;
    \If {misunderstanding user's interactive intent}
    \State \textbf{goto} \emph{loop};
    \Else
    \State executing the current task and detect interruption in real time;
    \If {no interruption}
    \State waiting the end of current task and \textbf{goto} \emph{loop};
    \Else
    \If {new command is stoping}
    \State forget the breakpoint and \textbf{goto} \emph{loop};
    \EndIf
    \If {new command has lower priority}
    \State interruption is invalid and current task will be continued to execute;
    \EndIf
    \If {new command has higher priority}
    \State interruption is valid and current task will be suspended;
    \EndIf
    \EndIf
    \EndIf
  \end{algorithmic}
\end{algorithm}

\section{Experimental Results and Analysis}

\subsection{Experimental Setup}

We completed real-life human-robot interaction experiments in the hall including eight interactions between a human and the robot mentioned by Table.~\ref{tab:Interction}. In our experiments, the parameters of human activity recognition model were set in \cite{Li2017Sequential}. In order to verify the effectiveness of adding human-mimicking decision mechanism for HRI system, we chose two simple task-based interactive mode, Interaction 5 and 7 shown in Table.~\ref{tab:Interction} respectively. We defined that the task of moving back has higher priority than circling, and set the speed of the chassis as 0.2 m/s and the moving distance as 4 m, the center of the circle trajectory was set as the center of the chassis and the robot spins around 20 laps. First, the user executed drawing circle to tell the robot circling, the robot would rotate around itself. Then assuming that the task of moving backwards needed to do immediately, so the user lifted his left hand to interrupt current circling and performed waving forwards. After a while, we repeat aforementioned two activities alternately and observe the responses of the robot. Here we recorded the experimental results using screen recording mode in first view and video recording in third view.

\subsection{Experimental results and analysis}

\begin{figure}[!t]
 \centering
 \includegraphics[width=0.48\textwidth]{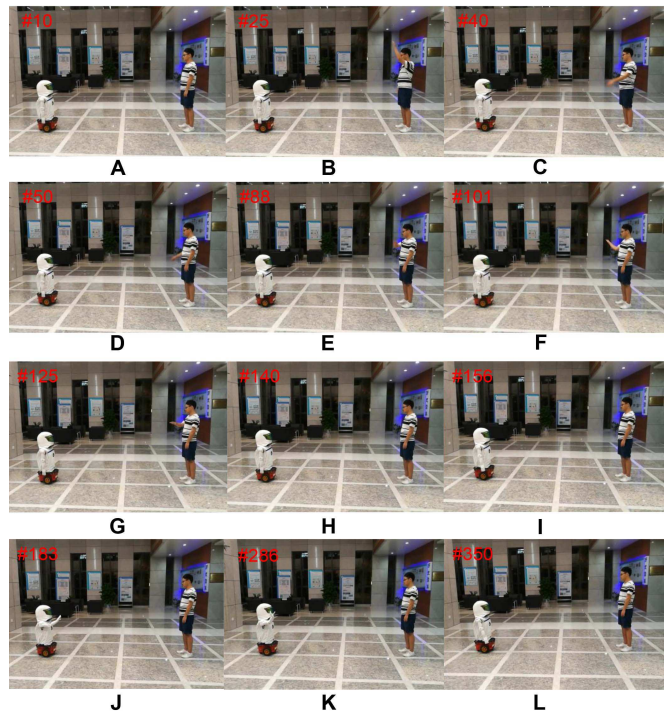}
 \caption{Handwaving interaction between robot and user.}
 \label{fig:real_life}
\end{figure}

Here, we take handwaving interaction for example. Fig.~\ref{fig:real_life} shows the results of the handwaving interaction. From A to C, the interaction switch was opened. From D to G, the user waved his right hand to greet the robot. From H to I, the robot analyzed the intention of user by calculating the output of LSTM network. From J to L, the robot responded user by waving its hand.

\begin{figure}[!t]
 \centering
 \includegraphics[width=0.48\textwidth]{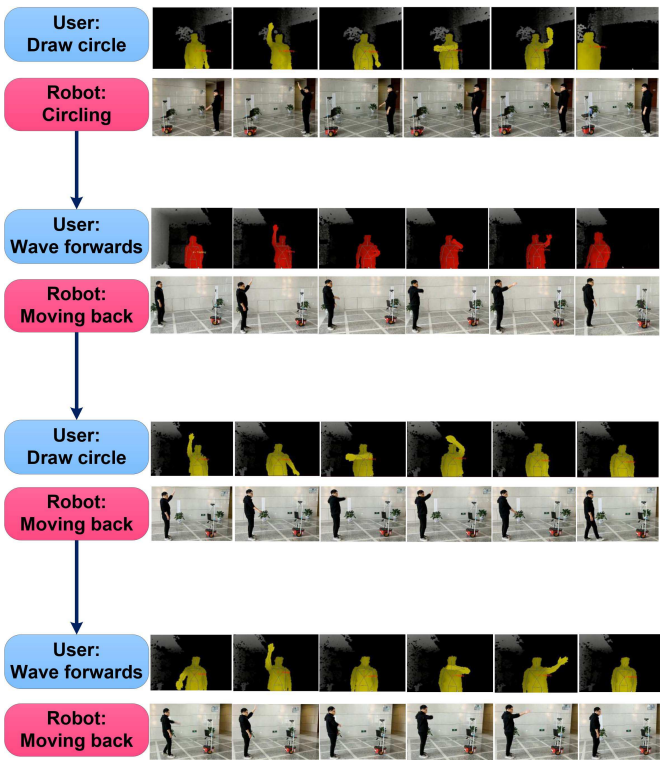}
 \caption{The verification results of human-mimicking decision mechanism for the HRI system.}
 \label{fig:decision}
\end{figure}

In the other experiment, in order to observe results more clearly, we chose a simplified robot platform without robotic arm because it does not use the robotic arm. Fig.~\ref{fig:decision} illustrates a series of results comprising of first view and third view. First, the user drew circle and the robot had a correct inferring. During the task, the user waved forwards, which has higher priority. The circling task is interrupted and the robot started moving back. After a while, the user tried to drew circle to interrupt the task of moving back, but failed because the circling has lower priority. When the user re-waved his hand forwards, the robot continued to move back from the breakpoint. From the result, we can see that the robot has become more smart with the ability to determine whether to interrupt current task according to the priority of the task.

\subsection{Discussion}

The excellent performance of real-life experiments had illustrated the effectiveness of our HRI system. Based on the framework, an increasing number of interactive activities can be designed to enrich the interactive ability of the robot. However, our HRI system can only be used indoors because the skeleton data was captured by Microsoft Kinect V1. I believe that it is a challenging task that how to accurately capture human skeleton data in outdoors in real time.

\section{Conclusion}

In this paper, we described our progress towards a HRI system for a service robot capable of recognizing several daily user activities and making human-mimicking decision. A practical and natural interactive switch based on filter model of attention was designed for the HRI system. The human-mimicking decision -- "priority should be given to important urgent matters" was also added into the HRI system. We presented results on several real-life interactive experiments, and demonstrated that our HRI system can realize excellent performance.

In the future, we will focus on the study of more intelligent interactive intent inferring for service robots. I believe that it is very interesting to make the robot more intelligent and smart.

\section*{Acknowledgment}

This work is partially supported by the National Natural Science Foundation of China under Grants 61673378 and 61421004.

\ifCLASSOPTIONcaptionsoff
  \newpage
\fi

\bibliographystyle{unsrt}
\bibliography{CYBER2018}

\end{document}